\documentclass{article}
\usepackage{epsf,spconf}
\usepackage{amsmath,amssymb,amsthm,epsfig}

\title {Blind Estimation of Multiple Carrier Frequency
Offsets}

\name {Yuanning Yu, Athina P. Petropulu, H. Vincent Poor$^+$ and
Visa Koivunen$^+$
\thanks{This work has been supported by NSF under grant Nos.
ANI-03-38807, CNS-06-25637 and CNS-04-35052, and by the Office of
Naval Research under Grant ONR-N-00014-07-1-0500.}}
\address {
Electrical \& Computer Engineering Department, Drexel University\\
$^+$School of Engineering and
Applied Science, Princeton University\\
  }

\begin{document}

\def\bW{\mbox{\boldmath $W$}}

\newcommand{\beq}{\begin{equation}}
\newcommand{\eeq}{\end{equation}}
\newcommand{\beqn}{\begin{eqnarray}}
\newcommand{\eeqn}{\end{eqnarray}}
\newcommand{\om}{\omega}

\sloppy

\ninept

\maketitle

\begin{abstract}

Multiple carrier-frequency offsets (CFO) arise in a distributed
antenna system, where data are transmitted simultaneously from
multiple antennas. In such systems the received signal contains
multiple CFOs due to mismatch between the local oscillators of
transmitters and receiver. This results in a time-varying rotation
of the data constellation, which needs to be compensated for at the
receiver before symbol recovery. This paper proposes a new approach
for blind CFO estimation and symbol recovery. The received base-band
signal is over-sampled, and its polyphase components are used to
formulate a virtual Multiple-Input Multiple-Output (MIMO) problem.
By applying blind MIMO system estimation techniques, the system
response is estimated and used to subsequently transform the
multiple CFOs estimation problem into many independent single CFO
estimation problems. Furthermore, an initial estimate of the CFO is
obtained from the phase of the MIMO system response. The Cramer-Rao
Lower bound is also derived, and the large sample performance of the
proposed estimator is compared to the bound.

\emph{keywords}-\textbf{Multi-user Systems, Distributed Antenna
Systems, Carrier Frequency Offset, Blind MIMO System
Identification}

\end{abstract}

\section{Introduction}

In any communication system, the received signal is corrupted by
carrier-frequency offsets (CFOs) due to the Doppler shift and/or
local oscillators drift. The CFO causes a frequency shift and a
time-varying rotation of the data symbols, which need to be
compensated for at the receiver before symbol recovery. This can be
achieved via pilot symbols. However, in the case of mobile systems
and rich scattering environments the effects of CFO become time
varying and even small errors in the CFO estimate tend to cause
large data recovery errors. This necessitates transmission of pilots
rather often, a process that lowers data throughput. In this paper
we deal with CFO estimation without the need for pilot symbols. In
single user systems, or in multi-antenna systems in which the
transmitters are physically connected to the same oscillator, there
is only one CFO that needs to be estimated. This is typically done
via a decision feedback Phase Lock Loop (PLL) at the receiver end.
The PLL is a closed-loop feedback control system that uses knowledge
of the transmitted constellation to adaptively track both the
frequency and phase offset between the equalized signal and the
known signal constellation. However, depending on the constellation
used during transmission, the PLL can have an M-fold symmetric
ambiguity, resulting in a limited CFO acquisition range, i.e.,
$|F_k|<1/(8T_s)$ for 4QAM signals, where $T_s$ is the symbol period.
 Moreover, the
PLL typically requires a long convergence time. Alternatively,
several methods have been proposed \cite{Ciblat}, \cite{Ghogho},
\cite{Gini}, \cite{Scott} \cite{Wang} that blindly estimate the CFOs
and recover the transmitted symbols using second-order cyclic
statistics of the over-sampled received signal. Blind CFO estimation
has also been studied in the context of orthogonal
frequency-division multiplexing (OFDM) systems, where the CFO
destroys the orthogonality between the carriers (see \cite{Koivunen}
and the references therein).

In a spatially distributed antenna system where data are transmitted
simultaneously from multiple antennas, the received signal contains
multiple CFOs, one for each transmit antenna. A PLL does not work in
that case as there is no single frequency to lock to. The literature
on estimation of multiple CFOs is rather sparse. Existing literature
on this topic focuses on pilot based CFO estimation. In
\cite{Frank}, the multiple CFOs were estimated by using pilots that
were uncorrelated between the different users. In \cite{CRBref1},
multiple CFOs were estimated via Maximum Likelihood based on
specially designed pilots. To account for multiple offsets,
\cite{Veronesi} proposed that multiple nodes transmit the same copy
of the data with an artificial delay at each node. The resulting
system was modeled as a convolutive single-input/single-output (SISO)
 system with time-varying
system response caused by the multiple CFOs. A minimum mean-square
error (MMSE) decision feedback equalizer was used to track and
equalize the channel and to recover the input data. Training symbols
were required in order to obtain a channel estimate, which was used
to initialize the equalizer.

Here we propose an approach for blind identification of multiple
CFOs and subsequent symbol recovery. The received base-band signal
is over-sampled, and its polyphase components are viewed as the
outputs of a virtual MIMO system. The time-varying contribution of
the CFOs, together with the transmitted symbols form the
multiple-input/multiple-output (MIMO) inputs, while the
time-invariant contribution of the CFOs along with fading channels
comprise the system response. By applying blind MIMO system
estimation techniques, the system response is estimated and used to
subsequently transform the multiple CFOs estimation problem into
many independent single CFO estimation problems. Furthermore, an
initial estimate of the CFO obtained from the phase of the MIMO
system response can be used to initialize the PLL eliminating the
symmetrical ambiguity problem.
 The resulting method has full acquisition range for
normalized CFOs, i.e., $|F_k|<P/(2T_s)$, where $P$ is the
over-sampling factor. To evaluate large sample performance we
establish the Cramer-Rao Bound (CRB) and compare the obtained mean
square error to it.

\section{System Model} \label{system}

We consider a distributed antenna system, where $K$ users transmit
simultaneously to a base station. Narrow-band transmission is
assumed here, where the channel between any user and the base
station is frequency non-selective. In addition, quasi-static
fading is assumed, i.e., the channel gains remain fixed during the
packet length. The continuous-time base-band received signal
$y(t)$ can be expressed as
\begin{equation}\label{cont-system}
 y(t)=\sum_{k=1}^K a_{k} x_k(t-\tau_k) e^{j2\pi F_k t} + w(t)
\end{equation}
where $a_k$ represents the effect of channel fading between the
$k-$th user and the base station and also contains the corresponding
phase offset; $\tau_k$ is the delay associated with the path between
the $k-$th user and the base station; $F_k$ is the frequency offset
of the $k-$th user and $w(t)$ represents noise; $x_k(t)$ denotes the
transmitted signal of user $k$: $x_k(t)=\sum_i s_k(i)p(t-iT_s)$
where $s_k(i)$ is the $i-$th symbol of user $k$; $T_s$ is the symbol
period; and $p(t)$ is a pulse function with support $[0,T_s]$.

Our objective is to obtain an estimate of
${\mathbf{s}}(i)=[s_1(i),...,s_K(i)]^T$ of the form

\begin{equation}\label{20}
{\hat {\bf s}} (i)={\tilde {\bf \Lambda} } {\bf P}^T{\bf s}(i)
\end{equation}
 where $\mathbf{P}$ is a column permutation
matrix and ${\tilde {\bf \Lambda}}$ is a constant diagonal matrix. These
are considered to be trivial ambiguities, and are typical in any
blind inference problem.

The received signal $y(t)$ is sampled at rate $1/T=P/T_s$, where the
over-sampling factor $P\ge K$ is an integer. In order to guarantee
that all the users' pulses overlap at the sampling times, the
over-sampling period should satisfy: $T_s/P \ge \tau_k, k=1,...K$,
which means that the over-sampling factor $P$ is upper bounded by
$T_s/min\{\tau_1,...,\tau_K\}$. Let $t=iT_s + m T, \quad
m=1,...,P-1$ denote the sampling times. The over-sampled signal can be
expressed as { \hspace{-1.5in}
\begin{eqnarray}
&&y_m(i)=y(iT_s+mT) \nonumber \\
&=& \sum_{k=1}^K a_k e^{j 2 \pi f_k
(iP+m)}x_k((i+\frac{m}{P})T_s-\tau_k) +w((i + \frac{m}{P} )T_s)
 \nonumber \\
&=& \sum_{k=1}^K a_{m,k} (s_k(i) e^{j 2 \pi f_k iP}) +w(i +
\frac{m}{P}), m=1,...,P-1\label{over-sampled}
\end{eqnarray}}
where $f_k=F_k T_s/P, \ (|f_k|\le 0.5)$ is the normalized frequency
offset between the $k-$th user and the base station, and the element
of the virtual MIMO channel matrix ${\mathbf A}$ is given as
\begin{equation}\label{virtual-channel}
a_{m,k}=a_k e^{j 2\pi mf_k} p(\frac{m}{P}T_s-\tau_k), \quad
m=1,...,P
\end{equation}
Defining ${\mathbf y}(i) \buildrel \triangle \over
=[y_{1}(i),...,y_{P}(i)]^T$; ${\mathbf A}=\{a_{m,k}\}$, a tall
matrix of dimension $P \times K $; ${\mathbf {\tilde s}}(i)
\buildrel \triangle \over=[s_1(i)e^{j 2 \pi f_1 i P},...,s_K(i)e^{j
2 \pi f_K i P}]^T$; and ${\mathbf w}(i)\buildrel \triangle
\over=[w(i+\frac{1}{P}),...,w(i+\frac{P}{P})]^T$, eq.
(\ref{over-sampled}) can be written in matrix form as
\begin{equation} \label{matrixform}
{\mathbf y}(i)={\mathbf A {\mathbf {\tilde s}}}(i) +{\mathbf w}(i)
\end{equation}

\section{Blind channel estimation and compensation of the CFOs}
\label{blind}

Let us make the following assumptions.
\begin{itemize}
\item
$\mathbf{A1})$ For each $m=1...P$, $w_{m}(.)$ is a zero-mean
Gaussian stationary random processes with variance $\sigma^{2}_{w}$,
and is independent of the inputs.

\item
$\mathbf{A2})$ For each $k$, $s_{k}(.)$ are a zero mean, independent
identically distributed (i.i.d.) stationary with nonzero kurtosis,
i.e., $\gamma_{s_k}^4= \mbox{Cum}[s_{k}(i), s^{*}_{k}(i), s_{k}(i),
s^{*}_{k}(i)]\ne 0.$ The $s_k$'s are mutually independent, we can
further assume that every user has unit transmission power, then
${\bf C}_{s} = {\bf I}$.

\item $\mathbf{A3})$ The over-sampling factor $P$ is no less than $K$.
\end{itemize}
Under assumption (A2), it is easy to verify that the rotated input
signals ${\tilde s}_k(.)$ are also zero mean, i.i.d, wide sense
stationary with nonzero kurtosis. Also, the ${\tilde s}_k(i)$'s are
mutually independent for different $k$'s. Assumption (A3) guarantees
that the virtual MIMO channel matrix $\mathbf A$ in
(\ref{matrixform}) has full rank with probability one. If the delays
of users are randomly distributed in the interval $[0,T_s/P)$, then
each row of the channel matrix can be viewed as drawn randomly from
a continuous distribution, thus the channel matrix has full rank
with probability one.

One can apply any blind source separation algorithm (e.g.,
\cite{JADE}) to obtain
\begin{equation}\label{parafacyield}
 {\hat{\bf A}} \buildrel \triangle \over = {\bf A}{\mathbf
P}{\mathbf\Lambda }
 \end{equation}
Subsequently, using a least-squares equalizer we can get an estimate
of the de-coupled signals ${\mathbf {\tilde s}}(i)$, within
permutation and diagonal scalar ambiguities as
\begin{equation}\label{channelest}
{\bf {\hat {\tilde s}}}(i)= {({\bf \hat{A}}^H \bf \hat{A})^{-1}{\bf
\hat{A}}}^H {\bf y}(i) = e^{jArg \{ {-\bf \Lambda} \}}{\bf |\Lambda|
}^{-1} {\bf P}^T{\bf \tilde s}(i)
\end{equation}
Denoting by $\theta_k$ the $k-$th diagonal element of $Arg\{ {\bf
\Lambda}\}$, the $k-$th separated input signal can be expressed as
\begin{equation}\label{decouple}
{\hat {\tilde s}}_k(i)=s_k(i)e^{j(-\theta_k+ 2 \pi f_k iP)}
\end{equation}
 Based on
(\ref{decouple}), any single CFO blind estimation method could be
applied to recover the input signal. Those methods can benefit by a
CFO estimate provided by the channel matrix estimate as follows. The
phase of the estimated channel matrix $ {\hat {\bf A}}$ equals
\begin{equation} \label{estCFO}
\Psi=Arg{\mathbf {\hat A}}= \left(%
\begin{array}{ccc}
 2\pi f_1 + \phi_1 & \ldots & 2\pi f_K + \phi_K \\
 \vdots & \ddots & \vdots \\
 2\pi f_1 P + \phi_1 & \ldots & 2\pi f_K P + \phi_K \\
\end{array}%
\right) {\mathbf P}
\end{equation}
where $\phi_k=Arg\{a_k\}+\theta_k$, which accounts for both the
phase of $a_k$ and the estimated phase ambiguity in
(\ref{decouple}). The least squares estimate of $f_k$ is given by
\begin{equation}\label{estCFO1}
{\hat f}_k=\frac{1}{2\pi}\frac{P (\sum_{p=1}^{P} p
\Psi_{p,k})-(\sum_{p=1}^P p)
(\sum_{p=1}^{P}\Psi_{p,k})}{P(\sum_{p=1}^{P}p^2)- (\sum_{p=1}^{P}
p)^2}
\end{equation}
We can write ${\hat f}_k = f_k+ \epsilon_k$ where $\epsilon_k$
represents the estimation error.

Noting that the de-coupled signals ${\hat {\tilde s}}_j(i)$ in
(\ref{decouple}) are shuffled in the same manner as the estimated
CFOs in (\ref{estCFO1}), we can use the estimated CFOs to compensate
for the effect of CFO in the decoupled signals (\ref{decouple}) and
get estimates of the input signals as
\begin{equation}\label{recovered}
{\hat {\bf s}}(i)=e^{jArg \{ {-\bf \Lambda} \}} {\bf P}^T{\bf s}(i)
\end{equation}

Due to the residual error in the estimated CFOs, we can only
compensate for a majority of the effect of CFO in (\ref{decouple})
and obtain
\begin{equation}\label{recovered-error}
{\hat s}_k(i)=s_k(i)e^{j(-\theta_k- 2 \pi \epsilon_k i P)}
\end{equation}

Let us apply the PLL to the recovered signals ${\hat s}_j(i)$ in
(\ref{recovered-error}), to further mitigate the effect of residuary
CFO $\epsilon_k $. For 4QAM signals, as long as $|P
\epsilon_k|<1/8$, it can be effectively removed by the PLL. Thus,
the CFO estimator (\ref{estCFO1}) can prevent the symmetric
ambiguity of the PLL, and can also greatly reduce the convergence
time of PLL. From (\ref{estCFO}), we can see that the CFO estimator
will achieve full acquisition range for the normalized CFO, i.e.,
$|f_k|<1/2$, which means we can deal with all continuous CFOs in the
range $F_k<P/(2T_s)$.

\section{Cramer-Rao Lower Bound}\label{secCRB}

To evaluate the large sample performance of the proposed method, we
establish the Cramer-Rao lower bound according to \cite{CRBref}. Via
central limit theory arguments, the received signals ${\bf y}$ can
be approximated as complex Gaussian signal, with zero mean, and
covariance matrix given by
\begin{equation}\label{covariance}
{\bf C}_y= {\bf AC}_{\tilde s}{\bf A}^H +\sigma_w^2 {\bf I}= {\bf
A}{\bf A}^H +\sigma^2_w {\bf I}
\end{equation}
The covariance matrix is valid under assumption A1) and A2). The
Gaussian assumption of the received signal is reasonable since the
received signal is a linear mixture of i.i.d. signals.

Let \begin{equation}\label{unknown} {\mathbf \alpha}= [{\bf f}^T,
{\mathbf{\rho}}^T, \sigma^2_w]^T
\end{equation}
where ${\bf f}^T=[f_1,...,f_K]^T$ is the vector of unknown CFOs, and
${\bf \rho}^T=[\tau_1,...,\tau_K]^T$ is the vector of random delays.
The parameter we are interested in is the CFOs ${\bf f}$, while
${\bf \rho}$ and ${\sigma^2_w}$ are the nuisance parameters.

Under the previous assumptions and the Gaussian approximation, the
Fisher Information Matrix (FIM) for the parameter vector ${\bf
\alpha}$ is given by \cite{CRBref}
\begin{equation}\label{FIM}
{\bf FIM}_{l,n}=T \mbox{Tr} ( \frac {\partial {\bf C}_y}{\partial
\alpha_l}{\bf C}_y^{-1}\frac {\partial {\bf C}_y}{\partial
\alpha_n}{\bf C}_y^{-1}), \quad l,n=1,...,2K+1
\end{equation}
Since we are only interested in the CFO parameter ${\bf f}$,
following the derivation in \cite{CRBref}, we can obtain that
\begin{equation}\label{CRB1}
\frac{1}{T}{\bf CRB}^{-1}({\bf f})={\bf G}^H {\bf G}- {\bf G}^H {\bf
\Delta}({\bf \Delta}^H {\bf \Delta})^{-1}{\bf \Delta}^H {\bf G} =
{\bf G}^H \Pi^\bot_\Delta {\bf G}
\end{equation}
where ${\bf G}$ and ${\bf \Delta}$ are defined as
\begin{equation}\nonumber
\frac{1}{T}{\bf FIM}=(\frac{\partial {\bf c}_y}{\partial {\bf
\alpha}^T})^H({\bf C}_y^T \otimes {\bf C}_y^{-1})(\frac{\partial
{\bf
c}_y}{\partial {\bf \alpha}^T}) = \left[%
\begin{array}{c}
 {\bf G}^H \\
 {\bf \Delta}^H \\
\end{array}%
\right]\left[%
\begin{array}{cc}
 {\bf G} & {\bf \Delta} \\
\end{array}%
\right]
\end{equation}
where ${\bf c}_y=\mbox{vec}({\bf C}_y)$ is a $P^2\times 1$ vector
constructed from columns of ${\bf C}_y$, and $\bf G$ is of dimension
$P^2 \times K$, while $\bf \Delta$ is of dimension $P^2 \times
(K+1)$.

To proceed, we just need to evaluate the derivatives of ${\bf c}_y $
with respect to $\bf \alpha$. First consider $\partial {\bf c}_y/
\partial {\bf f}^T$, it holds that
\begin{equation}\nonumber
\frac{\partial {\bf c}_y} {\partial f_k}=\mbox{vec}(\frac{\partial
{\bf C}_y}{\partial f_k})=\mbox{vec}([{\bf 0} \cdots {\bf d}_k
\cdots {\bf 0}]{\bf A}^H + {\bf A}[{\bf 0} \cdots {\bf d}_k^H \cdots
{\bf 0}]^T)
\end{equation}
with $ {\bf d}_k=\frac{j2\pi f_k}{P} ({\bf a}_k \odot [1,...,P]^T)$,
where $\odot$ is the Hadamard matrix product.

Similarly, we can get $\partial {\bf c}_y/
\partial {\bf \rho}^T$:
\begin{equation} \nonumber
\frac{\partial {\bf c}_y} {\partial
\tau_k}=\mbox{vec}(\frac{\partial {\bf C}_y}{\partial
\tau_k})=\mbox{vec}([{\bf 0} \cdots {\bf e}_k \cdots {\bf 0}]{\bf
A}^H + {\bf A}[{\bf 0} \cdots {\bf e}_k^H \cdots {\bf 0}]^T)
\end{equation}
with ${\bf e}_k= [e^{2\pi \frac{f_k}{P}} \frac{\partial
p(\frac{T_s}{P}-\tau_k)}{\tau_k},...,e^{2\pi f_k} \frac{\partial
p(T_s-\tau_k)}{\tau_k}]^T$.

Finally, we have that $\partial {\bf c}_y/\partial \sigma^2_w =
\mbox{vec} ({\bf C}_y^{-1})$. Now we have all the ingredients to
evaluate $CRB({\bf f})$ from (\ref{CRB1}).

\section{Simulation Results}\label{simulation}

In this section, we verify the validity of the proposed method via
simulations, under the following assumptions. The channel
coefficients $a_k, \quad k=1,...,K$ are zero-mean Gaussian random
variables. The waveform $p(.)$ used here is hamming window. The
continuous CFOs are randomly picked in the range
$[-\frac{1}{2T_s},\frac{1}{2T_s})$. The delays, $\tau_k$,
$k=1,...,K$ are uniformly distributed in the range of $[0,T_s/P)$.
The input signals used here are 4QAM signals. The estimation results
are averaged over 300 independent channels, and 20 Monte-Carlo runs
for each channel.

The blind source separation algorithm used is the JADE method, which
was downloaded from
http://www.tsi.enst.fr/~cardoso/guidesepsou.html.

\begin{figure}[htb4]
\begin{center}
\epsfxsize=3in \epsfbox{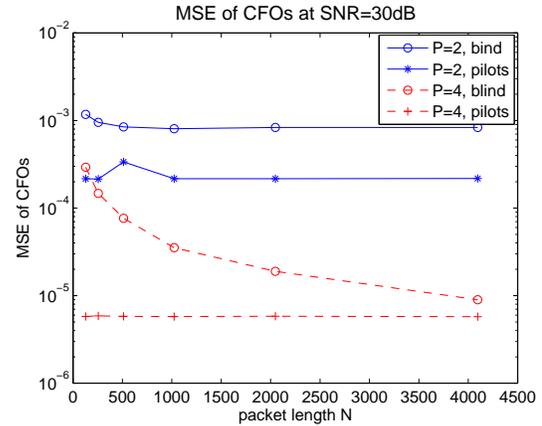} \\
\end{center}
\caption{MSE of CFOs vs $N$ for K=2, with SNR=30dB,
4QAM}\label{CFOMSE1}
\end{figure}

\begin{figure}[htb4]
\begin{center}
\epsfxsize=3in \epsfbox{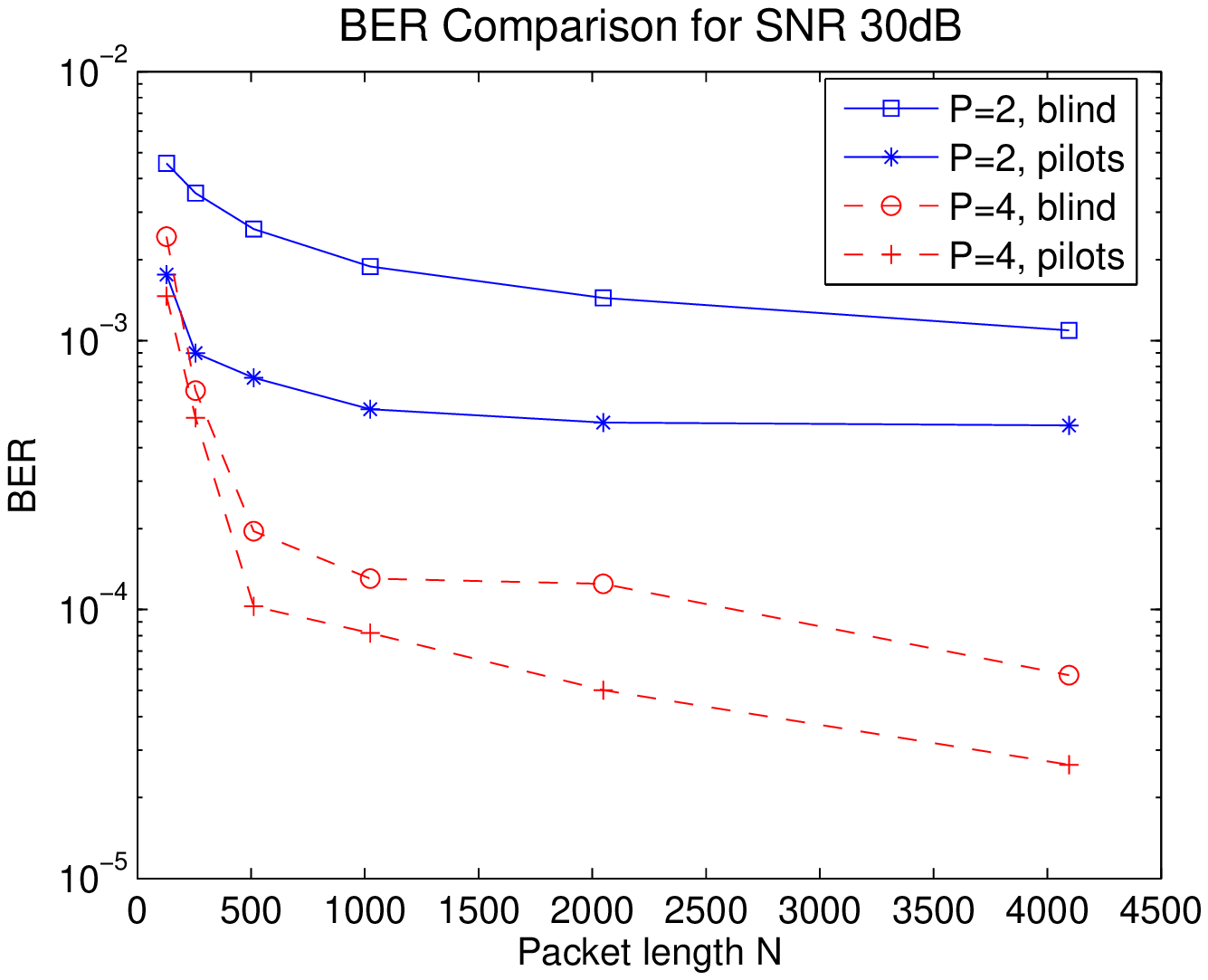} \\
\end{center}
\caption{BER vs $N$ for K=2, with SNR=30dB, 4QAM}\label{BER1}
\end{figure}

We show the performance of both the pilots-based method and the
proposed method at different data lengths and SNR set to 30dB. For
the pilots method, each user transmitted a pilot signal of length
$32$, and the pilots were random sequences uncorrelated between
different users. In Fig. \ref{CFOMSE1} we show the Mean Squares
Error (MSE) for the CFO estimator (\ref{estCFO1}) for different
values of the over-sampling factor $P$. To make the comparison fair
for different over-sampling factor $P$, the MSE is calculated based
on $\frac{1}{K}\sum_{k=1}^{K}[({\hat f}_k-f_k)P]^2 = \frac{1}{K}
\sum_{k=1}^{K}[({\hat F}_k-F_k)T_s]^2$. We can see that by
increasing $P$ we can get more accurate estimates of the CFOs. In
Fig. \ref{BER1}, we show the Bit Error Rate (BER) for different
$P$'s. For both the blind and the training based methods, the BER is
calculated based on the recovered signals after the PLL. As
expected, the BER performance also improves by increasing $P$. The
proposed method appears to work well even for short data length.

\begin{figure}[htb4]
\begin{center}
\epsfxsize=3in \epsfbox{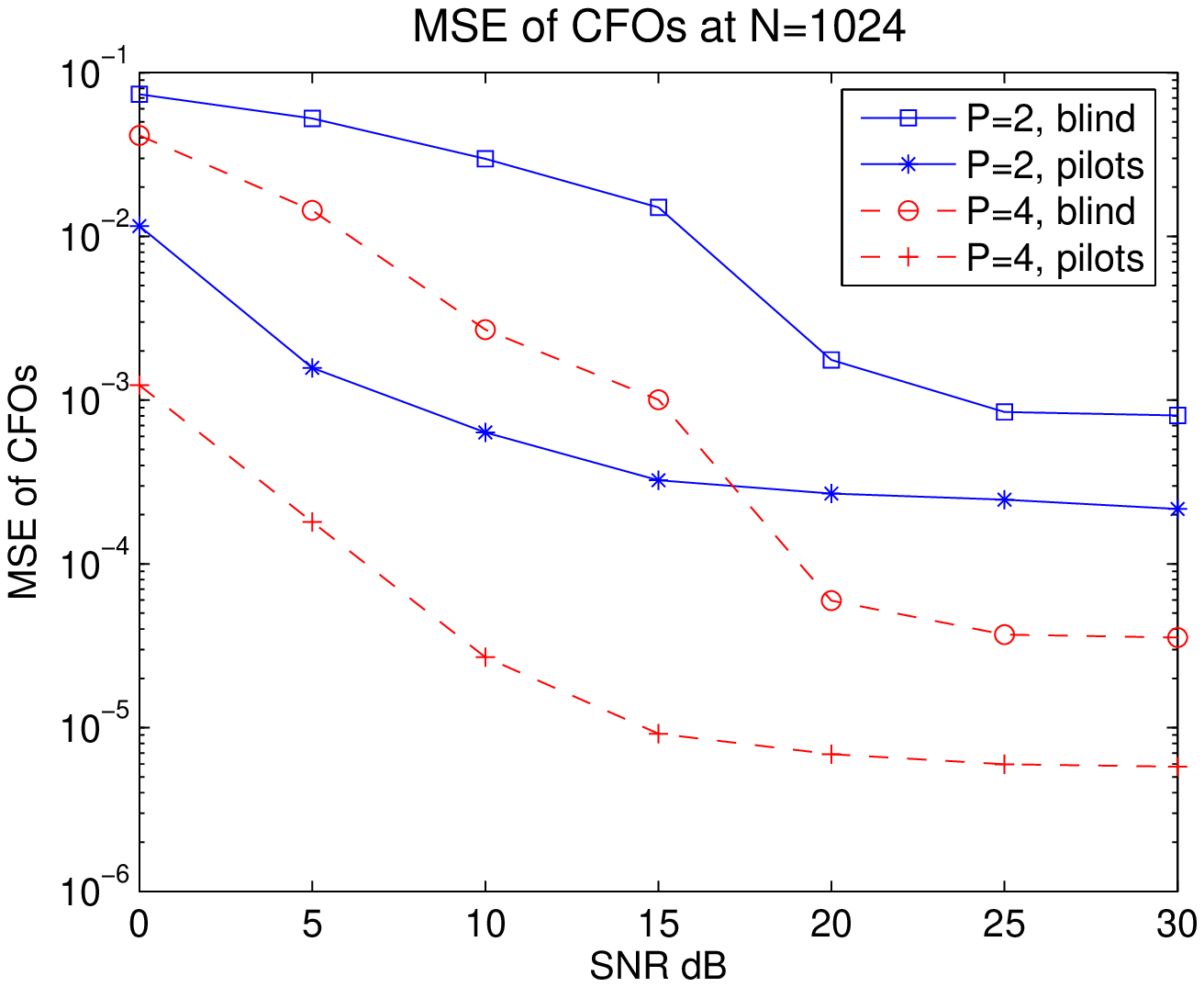} \\
\end{center}
\caption{MSE of CFOs vs SNR for K=2, 4QAM}\label{CFOMSE2}
\end{figure}

\begin{figure}[htb4]
\begin{center}
\epsfxsize=3in \epsfbox{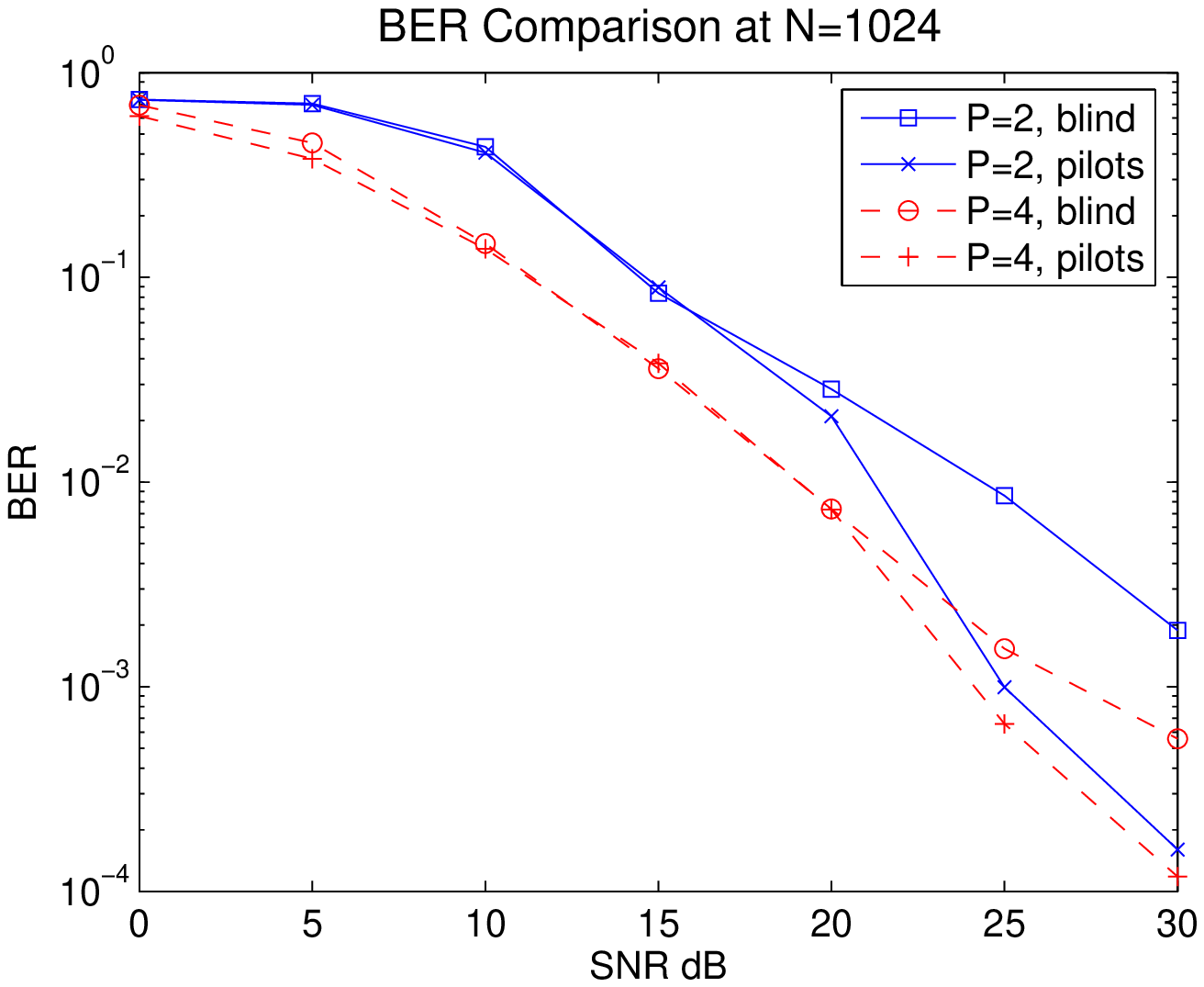} \\
\end{center}
\caption{BER vs SNR for K=2, 4QAM, T=1024}\label{BER2}
\end{figure}

Next we show the performance of both methods at various noise
levels. We set the packet length $N$ to $1024$. In Fig.
\ref{CFOMSE2}, we show the MSE of the blind CFO estimator
(\ref{estCFO1}) as well as the training based method. We can see
that by increasing $P$ we can get more accurate estimates of the
CFOs. In Fig. \ref{BER2}, we show the BER performance after PLL for
both blind and training based methods. We can see that the proposed
blind method has almost the same performance to the training based
method for SNR lower than 20dB, while the training based method can
achieve better BER performance for high SNR.

\begin{figure}[htb4]
\begin{center}
\epsfxsize=3in \epsfbox{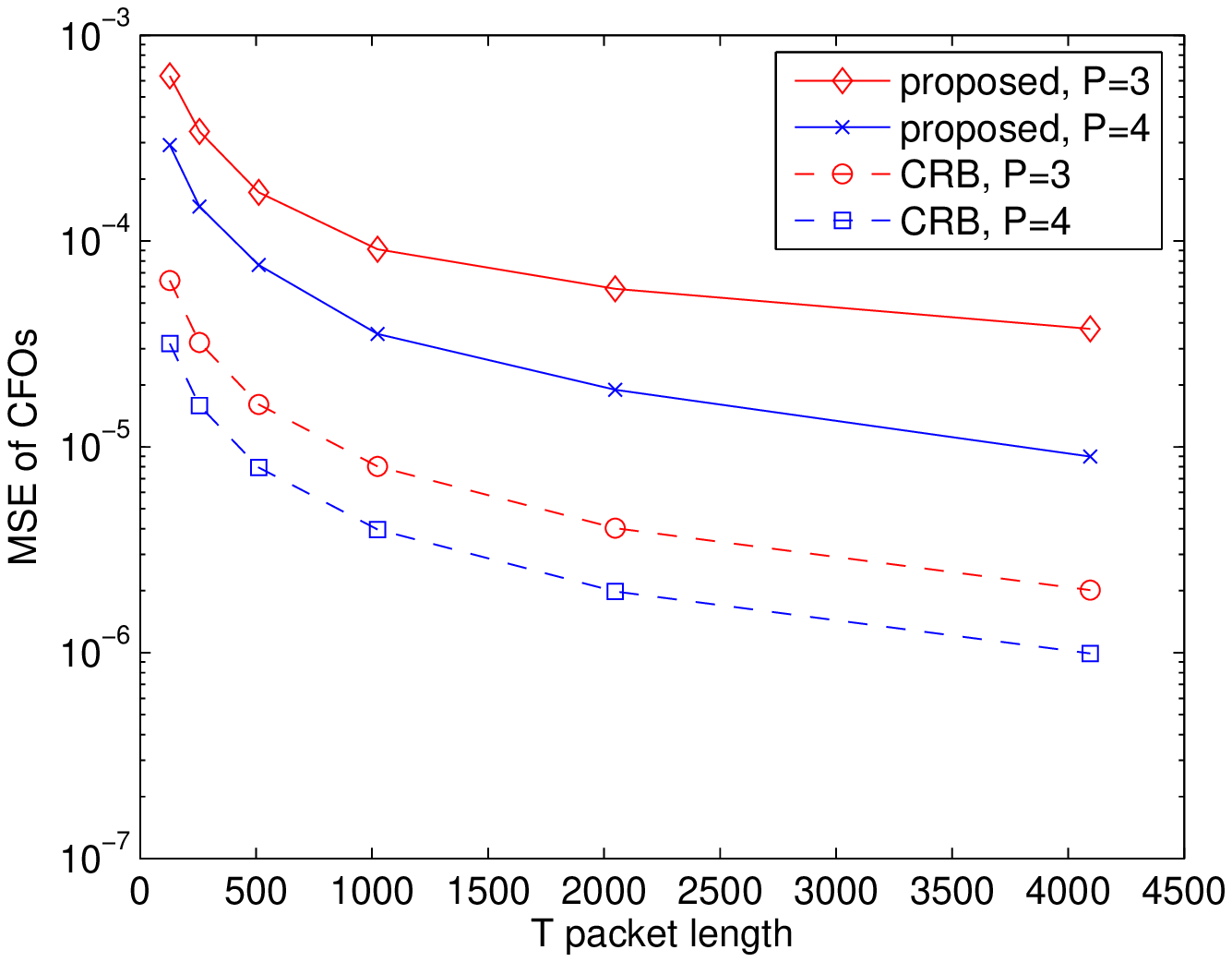} \\
\end{center}
\caption{ MSE of the CFOs for SNR=30 dB}\label{CRBfig1}
\end{figure}

The mean square error of the CFO estimator of (\ref{estCFO1}) is
plotted against the stochastic CRB derived in Section \ref{secCRB}.
In fig. \ref{CRBfig1}, we plotted the MSE of the CFOs, as well as
the CRB, as a function of the packet length $T$. We can see that the
MSE curves are parallel to the CRB, and no error floor presented in
the plot. Hence there is no bias in the estimates and the gap is
only due to excess variance in the estimates. One possible reason
for the existence of the excess variance is that we assume that we
know the exact channel structure in the derivation of the CRB, i.e.,
the waveform used in transmission, which reduces the number of the
unknown parameters. In the simulations, however, we did not assume
any extra knowledge of the channel structure.

We should note  that the PLL is important for good symbol recovery.
For example, without the PLL, even if the residual error
$P\epsilon_k={\hat F}_k-F_k$ is only $0.001$, the constellation will
be rotated to a wrong position after $0.25/0.001=250$ samples for
4QAM signals. To make sure that the PLL does not have the
symmetrical ambiguity, we need to guarantee that
$|P\epsilon_k|=|({\hat F}_k-F_k)T_s|<1/8$ for 4QAM transmission.
Thus, on the average, the maximum tolerable MSE for the CFO is in
the order of $10^{-2}$. From the simulations, we can see that the
achieved compensation is sufficient for practical systems and
commonly used modulation schemes.

\section{Conclusion}\label{conclusions}

In this paper we have proposed a novel blind approach for
identification of a distributed multiuser antenna system with
multiple CFOs. By over-sampling the received base-band signal, we
have converted the mulplte-input/single-output (MISO) problem into a
MIMO one. Blind MIMO system estimation yields the system response,
and MIMO input recovery yields the decoupled transmitted signals,
each one containing a CFO. By exploring the structure of the MIMO
systems response we can obtain a coarse estimate of the CFOs, which
can be combined with a decision feedback PLL to compensate for the
CFOs in the decoupled transmitted signals. The proposed blind method
has full acquisition range for normalized CFOs. We have provided a
Cramer-Rao bound (CRB) for the proposed blind CFO estimators. The
analytical results have been validated via simulations.

%\hspace{1.in}

%\hspace{1.in}

%\hspace{1.in}

%\begin{figure}[htb4]
%\begin{center}
%\epsfxsize=3.in \epsfbox{CRBT1024.eps} \\
%\end{center}
%\caption{MSE of the CFOs for T=1024}\label{CRBfig2}
%\end{figure}

\end{document}